\documentclass[conference]{IEEEtran}
\DeclareUnicodeCharacter{211D}{\,}
\usepackage[T1]{fontenc}    
\usepackage{hyperref}       
\usepackage{url}            
\usepackage{booktabs}       
\usepackage{amsfonts}       
\usepackage{nicefrac}       
\usepackage{microtype}      
\usepackage{lipsum}

\usepackage{filecontents}
\usepackage{comment}
\usepackage{enumerate}
\usepackage{amsfonts}
\usepackage{amssymb}
\usepackage{verbatim}
\usepackage{setspace}
\usepackage{mathrsfs}
\usepackage{tikz}
\usetikzlibrary{arrows,shapes}
\usepackage{graphicx}

\usepackage{subcaption}

\usepackage{etoolbox}

\usepackage{listings}
\lstdefinelanguage{encryption}
 {morekeywords=X, 
  otherkeywords={1,2,3,4,5,6,7,8,9,0,a,b,c,d,e,f},
  }

\usepackage{todonotes}
\usepackage{bm} 
\usepackage{float}
\usepackage{graphicx}
\usepackage{tikz}
\usetikzlibrary{shapes.geometric}
\usetikzlibrary{arrows}
\usetikzlibrary{arrows.meta,calc,decorations.markings,math,arrows.meta}

\usepackage{amsmath}
\usepackage{mathdots}
\usepackage{diagbox}
\usepackage{amsfonts}
\usepackage{amssymb}
\usepackage[style=ieee,mincitenames=1,maxcitenames=2,maxbibnames=10,doi=false,isbn=false,url=false,eprint=false]{biblatex}
\DeclareSourcemap{
  \maps[datatype=bibtex, overwrite]{
    \map{
      \step[fieldset=address, null]
      \step[fieldset=location, null]
    }
  }
}

\ifCLASSINFOpdf

\else

\fi

\DeclareRobustCommand*{\IEEEauthorrefmark}[1]{%
\raisebox{0pt}[0pt][0pt]{\textsuperscript{\footnotesize\ensuremath{#1}}}}

\begin{document}
%
\title{Advancing Blockchain Scalability: An Introduction to Layer 1 and Layer 2 Solutions}

\author{
\IEEEauthorblockN{ 
Han Song\IEEEauthorrefmark{1},
Zhongche Qu\IEEEauthorrefmark{2},
Yihao Wei\IEEEauthorrefmark{3},
}
\IEEEauthorblockA{\IEEEauthorrefmark{1} University of Southern California, Los Angeles, US}
\IEEEauthorblockA{\IEEEauthorrefmark{2} Columbia University, New York, US}
\IEEEauthorblockA{\IEEEauthorrefmark{3} University of Illinois Urbana-Champaign, Champaign, US}

\IEEEauthorblockA{hsong427@usc.edu\IEEEauthorrefmark{1}, 
zq2172@columbia.edu\IEEEauthorrefmark{2},
yihaow4@illinois.edu\IEEEauthorrefmark{3},  }
}

\maketitle

\begin{abstract}
Bitcoin’s rise has put blockchain technology into the mainstream, amplifying its potential and broad utility. While Bitcoin has become incredibly famous, its transaction rate has not match such a corresponding increase. It still takes approximately 10 minutes to mine a block and add it to the chain. This limitation highlights the importance of seeking scale-up solutions that solve the low throughput transaction rates. Blockchain's consensus mechanisms make peer-to-peer transactions becomes feasible and effectively eliminate the need for centralized control. However, the decentralized systems also causes a lower speed and throughput compared to centralized networks as we mentioned Bitcoin's block creation rates.  Two mainstreams scale-up solutions, Layer 1 scale-up and Layer 2 scale-up have been  implemented to address these issues.  Layer 1 level scalability enhancements happen at where traditional blockchain operates. This paper provides a deep examination of the components of the Layer 1 protocol and the scale-up methods that directly improve the lower level blockchain. We also address that Layer 1 solutions encounter inherent limitations although improvements were applied due to layer 1 storage costs and latency are high.
In addition, we discuss layer 2 protocols, advanced scalability techniques,  that elevate blockchain performance by handling transactions off the mainnet.  Our findings indicate that Layer 2 protocols, with their various implementations such as rollups and channels, significantly outperform Layer 1 solutions in terms of transaction throughput and efficiency. This paper discusses these Layer 2 scaling methods in detail, aiming to provide readers with a comprehensive understanding of these protocols and the underlying logic that drives their effectiveness. 

Keywords

Cryptography, Blockchain, Scalability, Web3
\end{abstract}

\section{Introduction}
Since Bitcoin successfully launched peer-to-peer transaction system without any third party involvement\cite{nakamoto2008bitcoin}, blockchain technology is brought to public and achieves fully decentralization . This success is attributed to blockchain's consensus mechanism by allowing anyone to verify transactions by hosting a blockchain node. Moreover, consensus mechanism sacrifices limitations, including low speed and throughput, compare to centralized system where single server handles requests across the whole network. For instance, Bitcoin adds a new mined block to the chain every 10 minutes because the transactions inside the block should be verified by all nodes on the blockchain, whereas traditional centralized payment processors can handle thousands of transactions per second.

As blockchain technology evolves and digital currencies gain wider acceptance, these limitations become increasingly problematic. High latency and low throughput are particularly evident when large numbers of people use Ethereum for transactions, leading to network congestion, delays, and higher gas fees\cite{bez2019scalability}. Therefore, finding scalable solutions is important to remediate the current issue and improve the further user experience.

This paper aims to elucidate scalability solutions in both Layer 1 and Layer 2 blockchains. First, we will examine Layer 1 blockchains and the scalability adjustments made to the consensus mechanism within this traditional blockchain framework. Following this, we will explore Layer 2 scalability, which encompasses various advanced strategies. We will discuss in detail Layer 2 strategies such as rollups\cite{thibault2022blockchain}, channels, and sidechains, and how they handle off-chain transactions while maintaining interactions with the Layer 1 blockchain. In this paper, we provide an in-depth knowledge sharing of the scalability challenges and solutions to solve in both layer 1 and layer 2 blockchains.

\section{Layer 1 Blockchain}
In the realm of blockchain technology, Layer 1 blockchain refers to the base networks that form the primary consensus mechanisms to accomplish the goal of  decentralization, such as Bitcoin, Ethereum, and Solana. These Layer 1 protocols ensures that basic on-chain activities such as send and receive transactions can be executed and interacted with consistency as no centralized agents handle and validate transactions. Bitcoin, as the first created Layer 1 blockchain, introduced the decentralized ledger system that made the protocols become real. Ethereum further enhanced this decentralized system with EVM and smart contract which support higher level interations with decentralized applications (dApps). Solana, a more recent Layer 1 blockchain, is built to solve layer 1 scalability challenges with high throughput and low latency. Layer 1 blockchains are essential for providing the basic infrastructure to support decentralized interactions.

\subsection{Consensus mechanism}
Consensus mechanism is one of the fundamental technologies in blockchain and it is the key to a decentralised system. In the absence of a central authority to verify the authenticity of transactions, blockchain networks employ consensus mechanisms at the Layer 1 level to ensure the validation of all transactions. Each blockchain network adopts a distinct consensus model, with Bitcoin utilizing Proof-of-Work (PoW) and Ethereum employing Proof-of-Stake (PoS)\cite{bach2018comparative}. These models enable individuals to participate in the transaction validation process, provided they meet specific requirements. Participants are rewarded with digital currencies for their honest verification work to the blockchain. Thus, consensus mechanisms achieve secure and transaction processing without third party involvement by the design instead rely on single party but incentivize all to participate in the blockchain ecosystem.

\subsection{Data Structure}
\begin{figure}[htp]  
\centering
\includegraphics[width=0.9\columnwidth]{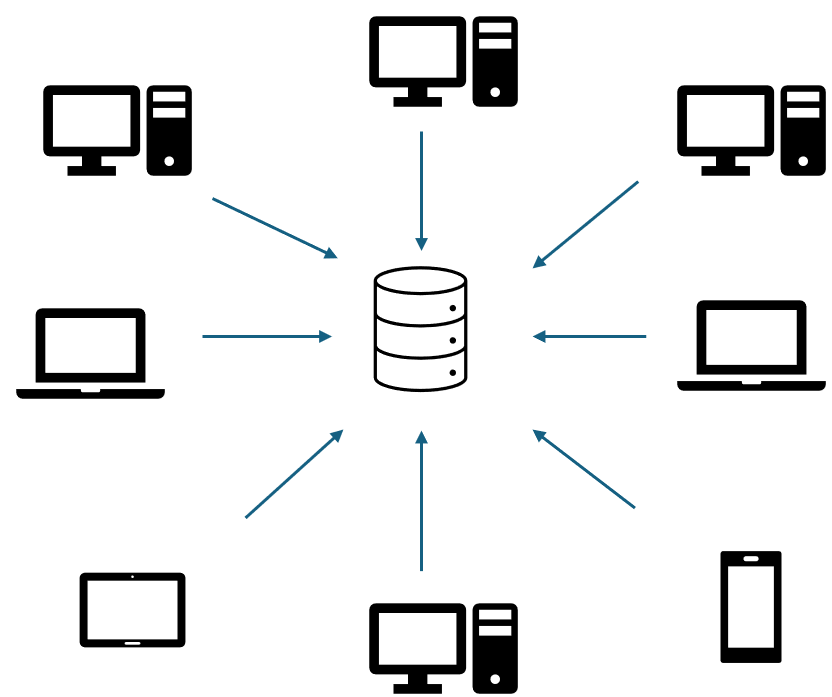}
\caption{Centralized database handles read and write requests from scattered clients}
\label{fig:central_db}
\end{figure}

Different to centralized database, Blockchain data storage sacrifices efficiency for decentralization. As depicted in Figure~\ref{fig:central_db}, In a centralized database, clients make requests directly to a single central entity and the centralized the entity is responsible for both data read and write. Centralized databases are good for transfer that highly leverage the read and write speed, such as Robotics and AI training~\cite{10503743,sekandi2023application,shi2023aging,shi2022pairwise, fudong:cikm22:cascade_vae,fudong:ecml23:storm,fudong:iccv23:mmst_vit}. On the other hand, in a decentralized system, data is stored on every Layer 1 blockchain network node. when new transaction is broadcasted over the blockchain, clients must wait for data synchronization to operate the write. Blockchain technology, referred to as a "Distributed Ledger,"\cite{wei2022survey} stores this ledger publicly across all Layer 1 network nodes. The ledger comprises a series of transactions that are bundled into blocks, with each newly created block linked to the preceding one, resembling a linked list. This structure ensures that all nodes maintain a consistent and transparent storage of transactions since everyone can trace the origin of each transaction in the network. Unfortunately, this data structure also experience high latency and time complexity compared to centralized databases.

\subsection{Fork}
\begin{figure}[htp]  
\centering
\includegraphics[width=1\columnwidth]{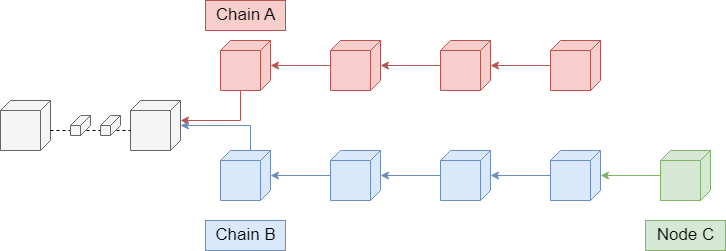}
\caption{After Node C adds the new mined block to the Chain B, Chain B becomes the longest chain and fork is resolved}
\label{fig:fork}
\end{figure}

Due to Blockchain's decentralized data structure, the data copies are not always consistent and nodes may keep different perspectives of the blockchain\cite{bazan2021bitcoin}. It is defined as Fork where different nodes may maintain different versions of the blockchain. Fork is a normal behavior and does not require heavy effort to resolve. Forks can be categorized into soft forks and hard forks. A soft fork is backward compatible and it occurs when some nodes have not upgraded to the latest protocol version but can still process on-chain transactions. In contrast, a hard fork is not backward compatible; hard fork requires protocol upgrade so that nodes can continue participate in on-chain activities. In Bitcoin, the node server always calculates the length of forked chains from the origin block and attach the latest mined block to the end of the longest forked chain. As illustrated in figure~\ref{fig:fork}, if two Bitcoin competing nodes, \textbf{Node A} and \textbf{Node B}, attempt to add a newly mined block to the existing chain simultaneously, the blockchain will fork into \textbf{Chain A} and \textbf{Chain B}. The fork may stay for a while as \textbf{Node A }and \textbf{Node B} always see their \textbf{Chain A} and \textbf{Chain B} the longest. Eventually, the fork will be resolved when another new block is mined by \textbf{Node C}. As the \textbf{Node C} must append this block to either \textbf{Chain A} or \textbf{Chain B}, thereby extending the chosen chain either \textbf{Chain A} or \textbf{Chain B} into the longest chain. Blocks from the shorter fork are re-queued for future processing.

\subsection{Cryptographic primitives}
Layer 1 blockchains employ cryptographic algorithms to prioritize security\cite{storublevtcev2019cryptography}. As we analyzed the blockchain data structure, transaction data are stored and shared on replicated nodes, It is crucial to keep system security while maintaining data transparency. Layer 1 blockchains achieve this balance through the use of asymmetric key cryptography. For example, in Bitcoin, users maintain a pair of keys: a private key and a public key. The public key is derived from the private key using a non-reversible cryptographic algorithm. User use the private key to sign the transaction and public key to send and receive the transactions. This key pair implemented by cryptographic algorithm therefore ensures secure data storage and user interactions on the Layer 1 blockchain.

\subsection{Scalability}

Layer 1 blockchains encounter high latency challenges when transaction volumes surge, leading to longer waiting times for users as their transactions are queued in a pool awaiting processing by miners. There are several strategies to overcome the scalability challenges. Increasing block size, allowing more transactions to be validated when creating the block, can improve transaction throughput. For instance, Bitcoin, which has a block size limit of 1 MB, can store approximately 2,000 transactions per block. Increasing the block size to 8 MB would theoretically enable the network to handle up to 16,000 transactions per block, thereby enhancing transaction throughput eight times.

Another strategy involves upgrading the consensus mechanism. Ethereum, for example, transitioned from the energy-intensive and time-consuming Proof-of-Work (PoW) consensus to Proof-of-Stake (PoS), which achieves more efficient transaction validation. Additionally, sharding has been proposed as a method to scale Layer 1 blockchains\cite{zhou2020solutions, hashim2022sharding, hong2022scaling}. Sharding divides the blockchain into smaller, manageable pieces, or shards, allowing transactions to be processed in parallel. Unlike traditional Layer 1 blockchains like Bitcoin, where each node maintains a complete copy of the blockchain and processes every transaction, sharding enables nodes to store and process only a portion of the data relevant to their shard. This division significantly boosts throughput by reducing the computational and storage burden on individual nodes, thus enhancing the overall efficiency of the blockchain network.

\section{Layer 2 Blockchain}
Blockchain is decentralized, immutable, append-only data structure where the blocks are stored dispersedly across all the nodes. This decentralized nature introduces scalability challenges, as opposed to centralized networks where a single server can be scaled directly by adding infrastructure such as memory and CPU. Consensus mechanism is at first priority to consider as essential upgrade in order to maintain the integrity of the blockchain. Layer 2 solutions have been developed as separate blockchains that inherit security and decentralization from Layer 1, aiming to increase transaction throughput and reduce transaction fees, which Layer 1 cannot efficiently achieve. In this section, we introduce Layer 2 blockchain protocols and what they can improve the scalability with their attributes, requirements and improvements.

\subsection{Rollups}

Rollups are a Layer 2 scaling solution that bundle multiple transactions into a single transaction. By processing transactions off-chain and sending the data back to the Layer 1 blockchain for consensus, rollups significantly enhance transaction throughput. There are two primary types of rollups: \textbf{Optimistic Rollups} and \textbf{Zero-Knowledge Rollups}.

Optimistic Rollups transfer computation and state storage off the Layer 1 Ethereum blockchain to reduce costs associated with computation and state changes. They batch multiple transactions together before submitting them to Ethereum Layer 1, compressing the data and sending it as calldata to lower data costs. Following a similar consensus mechanism to Ethereum Layer 1\cite{song2024unveiling}, Optimistic Rollups involve Layer 2 validators or sequencers who verify off-chain transactions by staking a fixed bond. The term "optimistic" reflects the assumption that all Layer 2 transactions are valid. After sequencers create a block for the batch of transactions, they submit it to the Layer 1 blockchain, where a fraud-proof mechanism allows for a fixed time window to challenge the results. If fraud is proven, the block is rejected, and the Layer 2 validator's stake is penalized.

\begin{figure}[htp]  
\centering
\includegraphics[width=0.9\columnwidth]{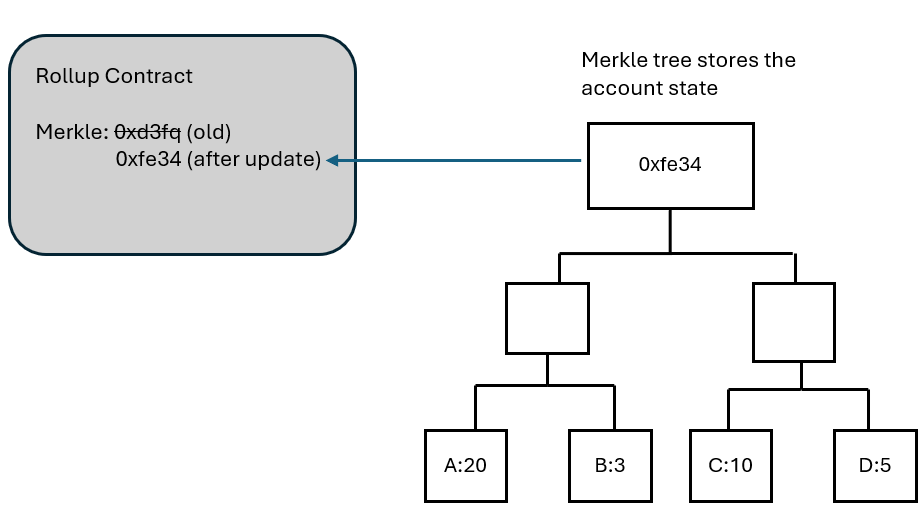}
\caption{post account state replaces the previous state in rollup contract}
\label{fig:op_merkle}
\end{figure}

Optimistic Rollups\cite{servaes2001roll, moosavi2023fast, pandey2021maintaining} interact with the Layer 1 blockchain via a set of smart contracts that store rollup blocks, monitor state updates, and track user deposits. As illustrated in Figure~\ref{fig:op_merkle}, the Merkle root is stored in the rollup contract for state updates. The previous state represents the old hash of the Merkle root, while the post-state is the new hash. Upon submission of the compressed batch transaction, the rollup contract verifies the previous state against the existing hashed state. If they match, the hashed state is updated to the post-state, finalizing the state changes. Notably, sequencers do not need to prove transaction validity when posting the batch transaction to Layer 1; instead, the fraud-proof mechanism initiates verification on the Layer 1 blockchain as described earlier.

\begin{figure}[htp]  
\centering
\includegraphics[width=1.0\columnwidth, height=2in]{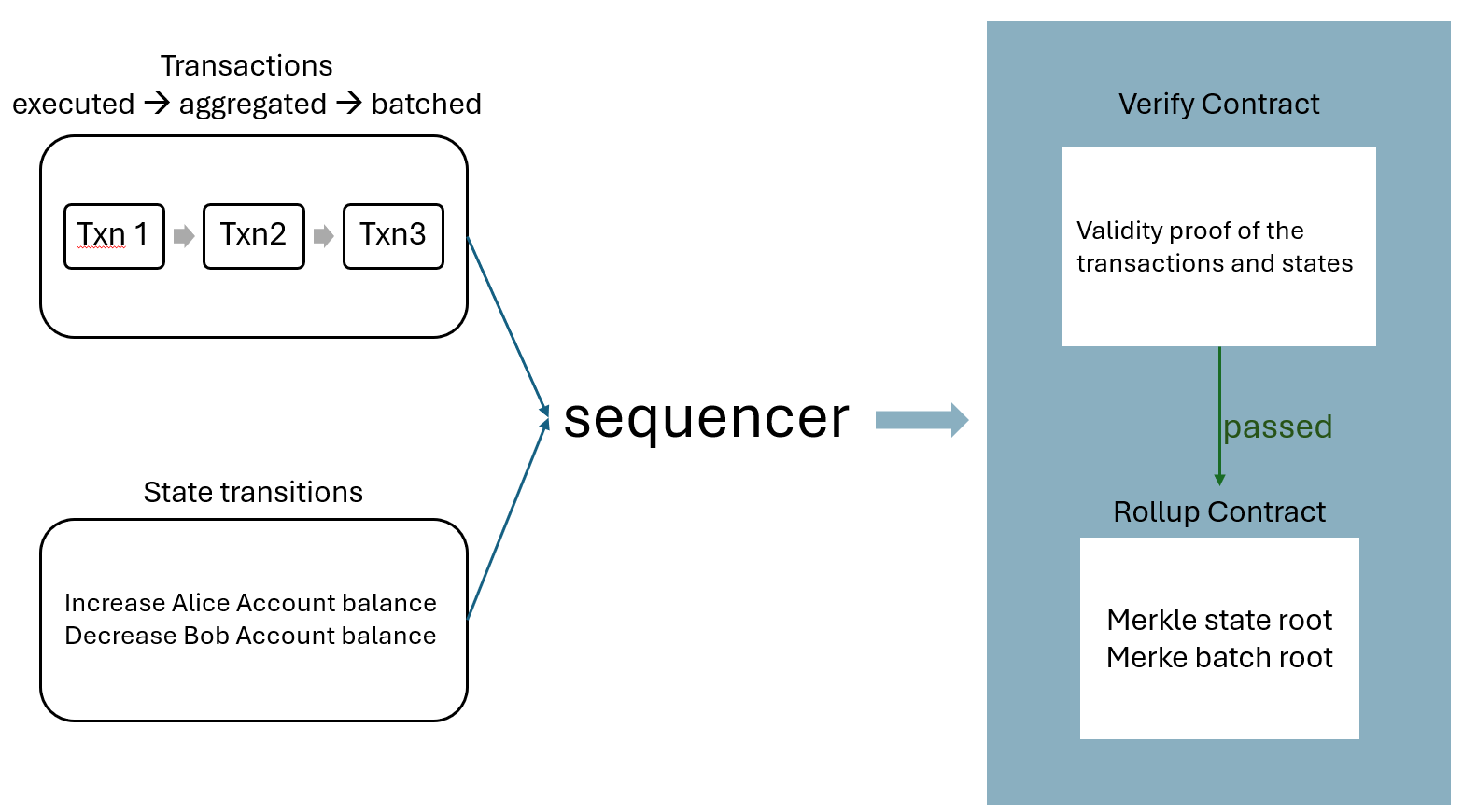}
\caption{workflow how sequencer interacts with on-chain contracts off-chain activities}
\label{fig:zk_flow}
\end{figure}

\begin{figure}[htp]  
\centering
\includegraphics[width=1.0\columnwidth, height=3in]{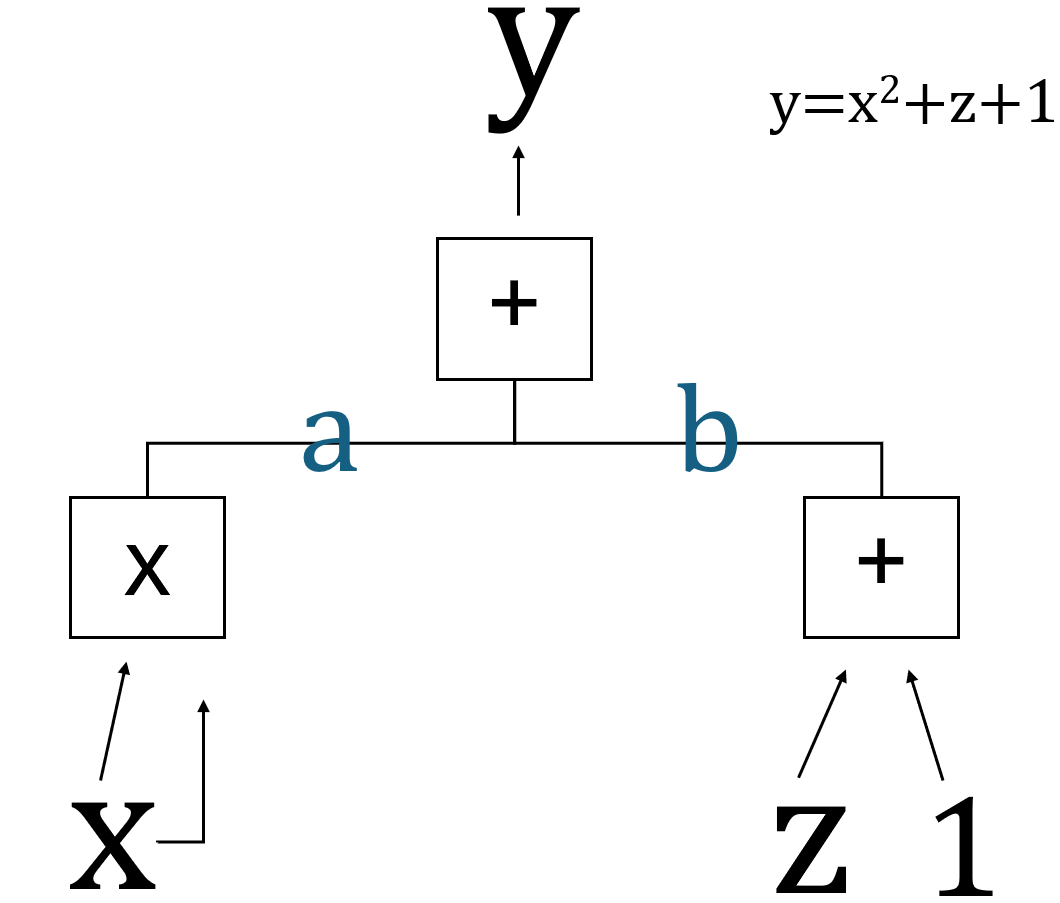}
\caption{arithmetic proving circuit shows the every step for calculation}
\label{fig:circuit}
\end{figure}

Zero-knowledge rollups(ZK-rollups) is another Layer 2 scaling on Ethereum\cite{sun2021survey, vcapko2022state,li2020privacy, zhou2024leveraging, tomaz2020preserving}. Similar to the optimistic rollups, ZK-rollups bundles transactions into batches and execute them off-chain in order to save storage and the computing cost. Instead of assuming the transactions are valid and posting all transaction data to layer 1 for later fraud-proof, ZK-rollups provide validity proof that assures the state updates are verified and ready to be finalized on Layer 1 blockchain. ZK-rollup protocol is controlled by smart contracts on Ethereum Layer 1. The main contract stores the rollup state root, transaction batch root, and a verifier contract is used to verify the validity proof. As illustrated in figure~\ref{fig:zk_flow}. When users on layer 2 blockchain submit transactions, the operator called sequencer starts the basic checks whether the sender and receiver has a valid state and enough funds. Sequencer then executes and aggregates those transactions into a batch follows with state transitions to their owner accounts, such as decreasing sender's balance and increasing the receiver's balance. The sequencer creates Merkle trees based on the batch transactions and state transitions to store in on-chain rollup contracts. However, the on-chain rollup contracts does not accept the updates automatically until a successful validity proof is produced and passed the verify contract. Validity proof is one of the most notable features in ZK-rollups. It confirms the correctness of the off-chain state transitions without revealing the transitons; in other word, it does not require the Layer 1 blockchain to re-execute the transactions from Layer 2 blockchain. Two types of zero-knowledge validity proof are used in ZK-rollups, \textbf{ZK-SNARK} (Zero-Knowledge Succinct Non-Interactive Argument of Knowdledge) and ZK-STARK (Zero-Knowledge Scalable Transparent Argument of Knowldege). They both serve as non-interactive proof which does not require third party authentication and avoids revealing the secret information. Non-interactive Zero-knowledge proof is achieved by arithmetic proving circuits. As depicted in figure~\ref{fig:circuit}, we explain how zero-knowledge proof work by implementing a simple calculation. Given a polynomial function \(y = x^2 + z + 1\), the prover needs to prove that \(x^2 + z + 1\) equals value \(y\) to verifier. In this case, we set \(x\) and \(y\) to public input and the verifier is not aware of any other information. The prover has to provide a proof that equation satisfied on both side without revealing the secret information. We usually refer the public input \(x\) and \(y\) as statement, secret input \(z\), intermediate variable \(a\), \(b\) as witness, and the intermediate calculations \(a = x * x\), \(b = z +1\), \(y = a + b\) are constraints. After the calculations are finished, the Zero-knowledge algorithm processes the circuits constraints to a VK(verification Key) that is similar to the commitment of the circuit. The verifier can verify the the proof through the VK provided. In this example, if we set \(x = 1, z = 2, y = 4\), the verifier knows prover makes solid proof with secret information and the equation satisfied when \(x = 1, y = 4\), without executing the arithmetic circuits. Layer 2 blockchain implements the validity proof in the same way that Zk-rollup operators compile the proving circuits includes Merkle proof for batch transactions, state roots, and account states. The proving circuits loop over every transactions to increase receiver's balance and decrease the sender's balance, update the account state and re-hash it with a new state root. After the proving circuits iterate all transactions, the last Merkle root becomes the latest state root. ZK-rollup sequencers now submit the validity proof to the Layer 1 on-chain verify contract for verification. The verify contract needs to check the Pre-state root equals Layer 2's last know state, post-state root equals the latest Layer 2 state root derived from proving circuit, and the batch root derived from batch transaction and hashed. If the proof is valid, the rollup contract will update its Merkle tree of the batch transaction root and reflect the changed Merkle state root. One of the benefits ZK-rollups provide is less delays for moving funds from Layer 2 blockchain to Layer 1 blockchain since the transaction executed once when verify contract approves the validity proof. However, the optimistic roll-up experiences huge delays for moving funds between different layers due to the fraud proof design that anyone challenges the transaction needs to re-execute the transaction. 

\subsection{Channels}

Channel is another Layer 2 protocol that used for blockchain scale-up\cite{li2020secure, papadis2020blockchain, negka2021blockchain, guo2023cross}. It allows users to make off-chain transactions between themselves and only post the final state to the Layer 1 blockchain and uses \textbf{multisig} contracts maintain the transactin validity. Channels are classified into two categories, Payment channel and State channel. In this section, we elucidate how channel benefits the blockchain scalability in layer 2.

Payment channel mainly focuses on the transactions where it enables instant payment transfers by allowing transactions to send and receive directly between participants. This functionality is also known as "two-way ledgers," To initiate a channel, participants (e.g., Party A and Party B) deposit funds into an on-chain smart contract. Once the channel is open, Party A and Party B can execute unlimited transactions(not exceeds the total amount they locked in the on-chain contract) with low latency and fees. The channel closes after the last transaction is finalized, and the participants must generate cryptographic signatures to validate the balance changes before they exit. The on-chain contract then updates their corresponding balances on the Layer 1 blockchain.

State channel aims to solve more than Payment channel is limited by enabling the tracking of state transitions. Similar to Payment Channels, State Channels involve a workflow for opening and closing the channel. To open a State Channel, participants deposit funds into an on-chain smart contract and cryptographically sign the initial state. These deposits can be used for payments and serve as a bond that ensures the participants won't commit any malicious actions; otherwise they are punished and their deposits are slashed. During the channel's operation, participants create and update states off-chain to achieve low-cost, high-speed transactions. In order to maintain the valid state transition and write it back to layer 1 blockchain, all participants must generate a cryptographic signature and acknowledge. When closing the channel, the on-chain smart contract verifies all transactions and state updates by using the previous participants' signatures. Depends on successful verification, the smart contract distributes the funds according to the participants' final account states and updates the Layer 1 blockchain accordingly.

\subsection{SideChain}

Sidechain is a layer 2 scale-up strategy which is designed to run as separate blockchains in parallel to the Layer 1 blockchain\cite{back2014enabling, gavzi2019proof, kiayias2020proof, yin2021sidechains, yang2020review,vispute2021scaling,tyagi2021study}.  Since sidechain is separate from Layer 1 blockchains, it utilizes different consensus algorithms, such as Proof of Authority, Delegated Proof of Stake, and Byzantine Fault Tolerance, instead of the original Ethereum consensus mechanism. Additionally, Sidechains aim to handle heavy computations and storage off-chain in the new blockchain. Thus it modify block parameters, such as increasing block size and block time, to enhance transaction speed and throughput. However, this alteration of the consensus mechanism potentially undermines the primary goal of consensus because hardware can become the bottleneck for running a full blockchain node with increased block size.

Each blockchain has its own ecosystem and environment, meaning assets on one blockchain cannot be directly used on another. This mirrors the real world scenario that US dollar is not spendable in China neither Chinese yuan can be spent in the US. If one who holds Chinese yuan wants to make transaction, he has to transfer the Chinese yuan to Us dollar from a bank. Blockchain bridge works as the bank role to connect and move asset between sidechain and Layer 1 blockchain. Bridges deploy smart contracts on both the sidechain and Layer 1 blockchain to verify and validate these transfers and ensure the transfers are valid. When bridge facilitate the asset transportation, the asset is not physically transferred. Bridge utilizes Two-way peg mechanism that lock the asset from original blockchain and mint equivalent amount of new asset on the target blockchain.

\section{Conclusion}
The scalability is a major problem on blockchain. There are a number of strategies addressed in both layer 1 and layer 2 blockchains. While Layer 1 blockchains can scale up by altering the consensus mechanism or change the inherent block size, they face the intrinsic high storage cost and latency. Layer 2 aims to improve the scalability and the main idea is to make transactions and computations off the layer 1 blockchain to reduce the cost, but also keep the blockchain's security and integrity. Rollups, through Optimistic and Zero-Knowledge methods, and Channels, via Payment and State mechanisms, demonstrate effective strategies for handling high transaction volumes and complex state transitions off-chain. we also discuss another layer 2 strategy, Sidechains that contribute by operating parallel to Layer 1 blockchains with alternative consensus algorithms to manage complex computations and storage independently. We discuss these various approaches of scalability in details and highlight the importance of continuous enhancement of blockchain scalability. We believe our study and explanations can help readers understand this domain better.




%
\newpage

\printbibliography

\end{document}